\documentclass[useAMS,usenatbib,referee]{mn2e}

\usepackage{natbib}
\usepackage{graphicx}
\usepackage{amssymb}
\usepackage{amsmath}
\usepackage{calc}
\usepackage{srcltx}

\title[DB white dwarfs: atmosphere electrical conductivity]
{Electrical conductivity of plasmas of DB white dwarf
atmospheres}

\author[V.~A.~Sre\'{c}kovi\'{c} et al]{V.~A.~Sre\'{c}kovi\'{c}$^{1}$,
Lj.~M.~Ignjatovi{\'c}$^{1,3}$ \thanks{E-mail: ljuba@ipb.ac.rs},
A.~A.~Mihajlov$^{1,3}$
\newauthor and M.~S.~Dimitrijevi\'{c}$^{2,3,4}$\\
$^{1}$Institute of Physics, P. O. Box 57, 11001 Belgrade,
Serbia\\
$^{2}$Astronomical Observatory, Volgina 7, 11160 Belgrade 74,
Serbia\\
$^{3}$Isaac Newton Institute of Chile, Yugoslavia Branch\\
$^{4}$Observatoire de Paris, 92195 Meudon Cedex, France
}

\begin{document}

\date{}

\pagerange{\pageref{firstpage}--\pageref{lastpage}} \pubyear{2009}

\maketitle

\label{firstpage}

\begin{abstract}

The static electrical conductivity of non-ideal, dense, partially
ionized helium plasma was calculated over a wide range of plasma
parameters:  temperatures $1\cdot 10^{4}\textrm{K} \lesssim T
\lesssim 1\cdot 10^{5}\textrm{K}$ and mass density $1 \times 10^{-6}
\textrm{g}/\textrm{cm}^{3} \lesssim \rho \lesssim 2
\textrm{g}/\textrm{cm}^{3}$. Calculations of  electrical
conductivity of plasma for the considered range of plasma parameters
are of interest for DB white dwarf atmospheres with effective
temperatures $1\cdot 10^{4}\textrm{K} \lesssim T_{eff} \lesssim
3\cdot 10^{4}\textrm{K}$.

Electrical conductivity of plasma was calculated by using the
modified random phase approximation and semiclassical method,
adapted for the case of dense, partially ionized plasma. The results
were compared with the unique existing experimental data, including
the results related to the region of dense plasmas. In spite of low
accuracy of the experimental data, the existing agreement with them
indicates that results obtained in this paper are correct.
\end{abstract}


\section{Introduction}

DB white dwarf atmospheres belong to the class of astrophysical
objects which have been investigated for a long time and from
various aspects \citep{bue70,koe80,sta93}. In the previous period,
the contribution of the authors of this paper was related to the
research of optical properties of DB white dwarf atmospheres within
the range of average effective temperatures $1\cdot 10^{4}\textrm{K}
\lesssim T_{eff} \lesssim 2\cdot 10^{4}\textrm{K}$. So, the papers
of \citet{mih92b} and of \citet{mih94,mih95} were dedicated to the
research of continual absorption in the optical part of EM spectra,
the paper of \citet{mih03b} - to the research of
chemi-ionization/recombination processes and of \citet{ign09} - to
the investigation of continual absorption in the VUV region of EM
spectra.

Recently, the transport properties of helium plasmas, characteristic
of some DB white dwarf atmospheres, attracted the authors'
attention, first of all the electrical conductivity. Namely, the
data on electrical conductivity of plasma of stars with a magnetic
field or moving in the magnetic field of the other component in a
binary system (see e.g. \citet{zha09}, \citet{pot09} and
\citet{rod09}) could be of significant interest, since they are
useful for the study of thermal evolution of such objects (cooling,
nuclear burning of accreted matter) and the investigation of their
magnetic fields. For example, \citet{kop70} and \citet{kop73}
studied electrical conductivity for stars of various spectral types,
in order to investigate the magnetohydrodynamic differences in their
atmospheres. Recently, \citet{maz07} investigated He conductivity in
cool white dwarf atmospheres, since the possibility of using these
stars for dating stellar populations has generated a renewed
interest in modelling their cooling rate \citep{fon01}. Also, the
transport processes occurring in the cores of white dwarfs (see e.g.
\citet{bai95} and numerous references therein) have been considered.
Moreover, electrical conductivity was particularly investigated for
solar plasma, since it is of interest for consideration of various
processes in the observed atmospheric layers, like the relation
between magnetic field and convection, the question of magnetic
field dissipation and the energy released by such processes (see e.
g. \citet{kop70} and references therein). For example \citet{fel93}
investigated the role of electrical conductivity in the construction
of a theoretical model of the upper Solar atmosphere, and
\citet{kaz06} considered the electrical conductivity of solar plasma
for magnetohydrodynamic simulations of the solar chromospheric
dynamo. Given that electrical conductivity plays an analogous role
in other stars as well, it is of interest to investigate its
significance, to adapt the methods for research into stellar plasma
conditions and to provide the needed data.

An additional interest for data on electrical conductivity in white
dwarf atmospheres may be stimulated by the search for extra-solar
planets. Namely \citet{jia98} have shown that a planetary core in
orbit around a white dwarf may reveal its presence through its
interaction with the magnetosphere of the white dwarf. Such an
interaction will generate electrical currents that will directly
heat the atmosphere near its magnetic poles. \citet{jia98} emphasize
that this heating may be detected within the optical wavelength
range as H$_\alpha$ emission. For investigation and modelling of
mentioned electrical currents, the data on electrical conductivity
in white dwarf atmospheres will be useful.

One of the most frequently used  approximations for consideration of
transport properties of different plasmas is the approximation of
"fully ionized plasma"
\citep{spi62,rad76,ada80,kur84,rop89,dju91,nur97,zai00,ess03}. It
was shown that the electrical conductivity of fully ionized plasmas
can be successfully calculated using the modified random-phase
approximation (RPA) \citep{dju91,ada94a,ada94b} in the region of
strong and moderate non-ideality, while the weakly non-ideal plasmas
were successfully treated within the semiclassical approximation
(SC) \citep{mih93,vit01}. In practice, even the plasmas with a
significant neutral component are treated as fully ionized in order
to simplify the considered problems,
\citep{rop89,ess98,zai00,ess03}. However, our preliminary estimates
have shown that such an approach is not applicable for the helium
plasmas of DB white dwarf atmospheres described in \citep{koe80},
where the influence of neutral component cannot be neglected.

Therefore, an adequate method for calculations of electrical
conductivity of dense, partially ionized helium plasmas is developed
in this paper. This method represents a generalization of methods
developed in \citet{dju91} and \citet{mih93}, namely modified RPA
and SC methods, and gives a possibility to estimate the real
contribution of the neutral component to the static electrical
conductivity of the considered helium plasmas within a wide range of
mass densities ($\rho$) and temperatures ($T$).

The calculations were performed for helium plasma in the state of
local thermodynamical equilibrium with given $\rho$ and $T$ for
$1\cdot 10^{4}\textrm{K} \lesssim T \lesssim 1\cdot
10^{5}\textrm{K}$ and $1 \times 10^{-6} \textrm{g}/\textrm{cm}^{3}
\lesssim \rho \lesssim 2 \textrm{g}/\textrm{cm}^{3}$. The obtained
results are compared with the corresponding experimental data
(\citet{min80,ter02,shi03}). For the calculations of plasma
characteristics of DB white dwarf atmospheres the data from
\citep{koe80} were used.

\section{Theory}

\subsection{The plasma electrical conductivity}

On the basis of the previous paper of \citet{ada80}, a modified RPA
method for calculation of static conductivity of fully ionized
plasma was developed in \citet{dju91} and \citet{ada94a}. The
principal role in this method belongs to the formula for
energy-dependent electron-electron (ee) and electron-ion (ei)
relaxation times $t_{ee;ei}(E) = t^{RPA}_{ee;ei}$, where $E$ is the
energy of single electron state, determined as a sum over the
Matsubara frequencies by using the methods of Green function theory.
This method is especially suitable for calculation of electrical
conductivity of dense non-ideal plasmas with electron density
($N_{e}$) larger than $10^{17}$ cm$^{-3}$. In the region $N_{e} <
10^{17}$ cm$^{-3}$ the static conductivity of fully ionized plasmas
can be determined well using the SC method developed in
\citet{mih93}, which is also based on electron relaxation time
$t_{ee;ei}(E) = t^{SC}_{ee;ei}$. It is important that SC method
gives practically the same results as RPA method in a wider region
of the electron densities around the value of $N_{e}=10^{17}$
cm$^{-3}$. SC method was tested from this aspect in \citet{vit01},
where it was experimentally verified through comparison with the
results from \citet{spi62} and \citet{kur84}, just for the helium
plasmas.

However, as it was already mentioned, the helium plasma of the
considered DB white dwarf atmospheres contains significant neutral
atom component, as follows from \citet{bue70} and \citet{koe80}.
Because of that, we will start here from the fact that in both RPA
and SC methods effective electron relaxation times $t_{ee;ei}(E)$
can be expressed as
\begin{equation}
\label{eq:tau_e} \frac{1}{t_{ee;ei}(E)} = \nu_{ee;ei}(E),
\end{equation}
where $E$ is the electron energy, and $\nu_{ee;ei}(E)$ - the
corresponding total electron-electron and electron-ion collision
frequency. This gives a possibility to generalize the modified RPA
and SC methods for the case of partially ionized plasmas, replacing
$1/t_{ee;ei}(E)$ by $\nu_{e;tot}(E)$
\begin{equation}
\label{eq:nu_tot} \nu_{e;tot}(E)= \nu_{ee;ei}(E) + \nu_{ea}(E) =
\frac{1}{t_{ee;ei}(E)} + \nu_{ea}(E),
\end{equation}
where $\nu_{ea}(E)$ is an effective electron-atom collision
frequency. Consequently, the basic RPA and SC expressions for the
static electrical conductivity $\sigma_{0}$ transform to the
corresponding Frost-like expressions
\begin{equation}
\label{eq:sigma0} \sigma_{0} = \frac{4 e}{3 m} \int_{0}^{\infty} E
\cdot w(E) \cdot \frac{1}{\displaystyle\left[\frac{1}{t_{ee;ei}(E)}+
\nu_{ea}(E)\right]}\cdot\frac{df_{FD}(E)}{dE}dE,
\end{equation}
where $m$ and $e$ are the mass and the modulus of charge of the
electron, $w(E)$ is the density of the single electron states in the
energy space, $f_{FD}(E) \equiv f_{FD}(E;T,N_{e})$ is the
Fermi-Dirac distribution function for given $N_{e}$ and temperature
$T$, and $t_{ee;ei}(E) = t^{RPA}_{ee;ei}$ or $t_{ee;ei}(E) =
t^{SC}_{ee;ei}$.

On the basis of composition and temperature data of the considered
DB white dwarf atmospheres \citep{bue70,koe80}, one can note that
only four components of these atmospheres are important for
determination of $\sigma_{0}$: free electrons, $He^{+}$ and
$He^{2+}$ ions, and helium atoms. In accordance with this, the
expression for $1/t^{RPA}_{ee;ei}$ from \citet{dju91} can be
presented in the form
\begin{equation}
\label{eq:tau_e1} \displaystyle \frac{1}{t^{RPA}_{ee;ei}(E)}=
\displaystyle \frac{4 \pi m N_{e} e^{4}k_{B}T}{(2 m E)^{3/2}} \int
_{0}^{\sqrt{8 m E}/\hbar} \frac{dq}{q} \sum_{n} \left\lbrace
\frac{Z_{e}^{2} \Pi_{e;n}(q)}{N_{e} \varepsilon_{n}^{3}(q)} +
\sum_{j=1,2}\frac{Z_{ij}^{2} \Pi_{ij;n}(q)}{N_{ij}
\varepsilon_{n}^{3}(q)} \right\rbrace,
\end{equation}
where $\hbar$ and $k_{B}$ are the Planck and Boltzmann constants
respectively, $Z_{e}\cdot e$ and $Z_{i1,i2} \cdot e$ - the charges
of electron and $He^{+}(1s)$ and $He^{2+}$ ions ($Z_{e} = -1$,
$Z_{i1} = 1$, $Z_{i2} = 2$), $N_{i1}$ and $N_{i2}$ - the
corresponding helium ion densities, $\Pi_{e;n}(q)$, $\Pi_{i1;n}(q)$
and $\Pi_{i2;n}(q)$ - the electron and ion polarization operators,
$\varepsilon_{n}(q)$ - the dielectric function, $n=0, \pm 1, \pm 2
...$, and summation is extended over all the Matsubara frequencies
$\Omega_{n}=2\pi n k_{B}T/\hbar$. The detailed expressions for
polarization operators and dielectric functions are given in
\citet{dju91} and \citet{ada94a}.

The SC expression for $\sigma_{0}$ is applied to the outer layers of
the considered DB white dwarf atmospheres where $N_{e} < 10^{17}$
cm$^{-3}$ and the presence of $He^{2+}$ ions can be neglected.
Because of that the expression for $1/t^{SC}_{ee;ei}$, in accordance
with \citet{mih93}, can be taken in the form
\begin{equation}
\label{eq:tau_e2} \displaystyle \frac{1}{t^{SC}_{ee;ei}(E)}=
\left[\frac{1}{\chi_{ee}} \frac{(2m)^{1/2}E^{3/2}}{2e^{4}Z_{i1}
N_{e}} \frac{1}{\ln(1+\Lambda_{i}^{2})^{1/2}}\right]^{-1} , \qquad
\Lambda_{i}=\frac{2E}{Z_{i1}e^{2}}\cdot r_{ci},
\end{equation}
where $k$ is Boltzmann constant and $Z_{i1}=1$. The correction
factor $1/\chi_{ee}=1/\chi_{ee}(Z_{i1},T)$ is determined within SC
method, while $r_{ci}\sim (4 \pi e^{2} N_{e}/kT)^{-1/2}$ is the
corresponding screening length which is an external parameter of the
theory. The values of $1/\chi_{ee}$ for $Z_{i1}=1$ and $10^{4}K \le
T \le 10^{5}K$ are taken from \citet{mih93}. Here, value
$r_{ci}=r_{n;i}$ is taken as the screening length, where the ion
neutrality radius $r_{n;i}$ is given by the expressions obtained in
\citet{mih09c}. Such a choice of the screening length provides
overlapping SC and RPA values of $\sigma_{0}$ in a wider region of
the electron densities around the value of $N_{e}=10^{17}$
cm$^{-3}$.

Finally, the electron-atom collision frequency $\nu_{ea}(E)$ in
Eq.~(\ref{eq:sigma0}) is given here by the known expression
\begin{equation}
\label{eq:nu_ea} \nu_{ea}(E)=N_{a}\cdot v(E) \cdot Q_{ea}^{tr}(E),
\end{equation}
where $N_{a}$ is the $He(1s^{2})$ atom density, $v(E)=(2E/m)^{1/2}$-
relative electron-atom velocity, and $Q_{ea}^{tr}(E)$- the transport
cross-section for the elastic $e-He(1s^{2})$ scattering
\citep{mot70}.

\subsection{The electron-atom transport cross-section}

The exact quantum-mechanical calculation of the transport
cross-section for elastic electron-atom collisions is a very hard
problem in itself. However, in the case of $e-He(1s^{2})$ scattering
the problem becomes easier within the considered temperature range
where all non-elastic collision processes can be neglected. Namely,
in this case it can be treated as scattering of electrons in an
adequately chosen model potential $U(r)$, where $r$ is the distance
between the electron and the nucleus of $He(1s^{2})$ atom. Thus, the
electron movement is described by a wave function
$\displaystyle{\Psi(r,\theta,\phi)=
\frac{\chi_{l}(r)}{r}Y_{lm}(\theta,\phi)}$, where
$Y_{lm}(\theta,\phi)$ is a spherical harmonic function of degree $l$
and order $m$, while the function $\chi_{l}(r)$ satisfies the radial
Schrodinger equation
\begin{equation}
\label{eq:RSE} \left[-\frac{1}{2}\frac{d^{2}r}{dr^{2}} + U(r) +
\frac{l(l+1)}{2r^{2}} \right]\chi (r) = E \chi (r),
\end{equation}
given in atomic units. Following the previous paper of
\citet{ign97}, we take the model potential $U(r)$ in the form
\begin{equation}
\label{eq:U(r)}
U(r) = \left\lbrace
\begin{array}{ll}
\displaystyle{ U_{0}(r) = - \frac{Z}{r} + \frac{q}{r + r_{0}} + \frac{Z - q}{r_{0}}, \qquad 0 < r < r_{i} }\\
\displaystyle{ U_{m}(r) = ar^{2} + br + c , \qquad  r_{i} < r < r_{f} } \\
\displaystyle{ U_{as}(r) = - \frac{\alpha}{2\left(r^{2} +
h^{2}\right)^{2}} \qquad  r_{f}  < r < \infty. }
\end{array}
\right.
\end{equation}
where $Z=2$ is the charge of the nucleus of a helium atom, $q$ and
$Z-q$ describe the redistribution of electrons in the $1s$ shell of
the helium atom ($0<q<2$), $\alpha$ is the polarizability of
$He(1s^{2})$ atom, and $r_{0}$ has the meaning of the average radius
of atom. Parameters $a$, $b$, $c$ and $h$ are determined from the
conditions of continuity and smoothness of the potential $U(r)$ at
the points $r_{i}$ and $r_{f}$ where $0<r_{i}<r_{0}$ and
$r_{f}>r_{0}$. The values of the mentioned quantities are given in
Table~(\ref{tab:koef_U}).

\begin{figure}
\centerline{\includegraphics[width=\columnwidth,
height=0.75\columnwidth]{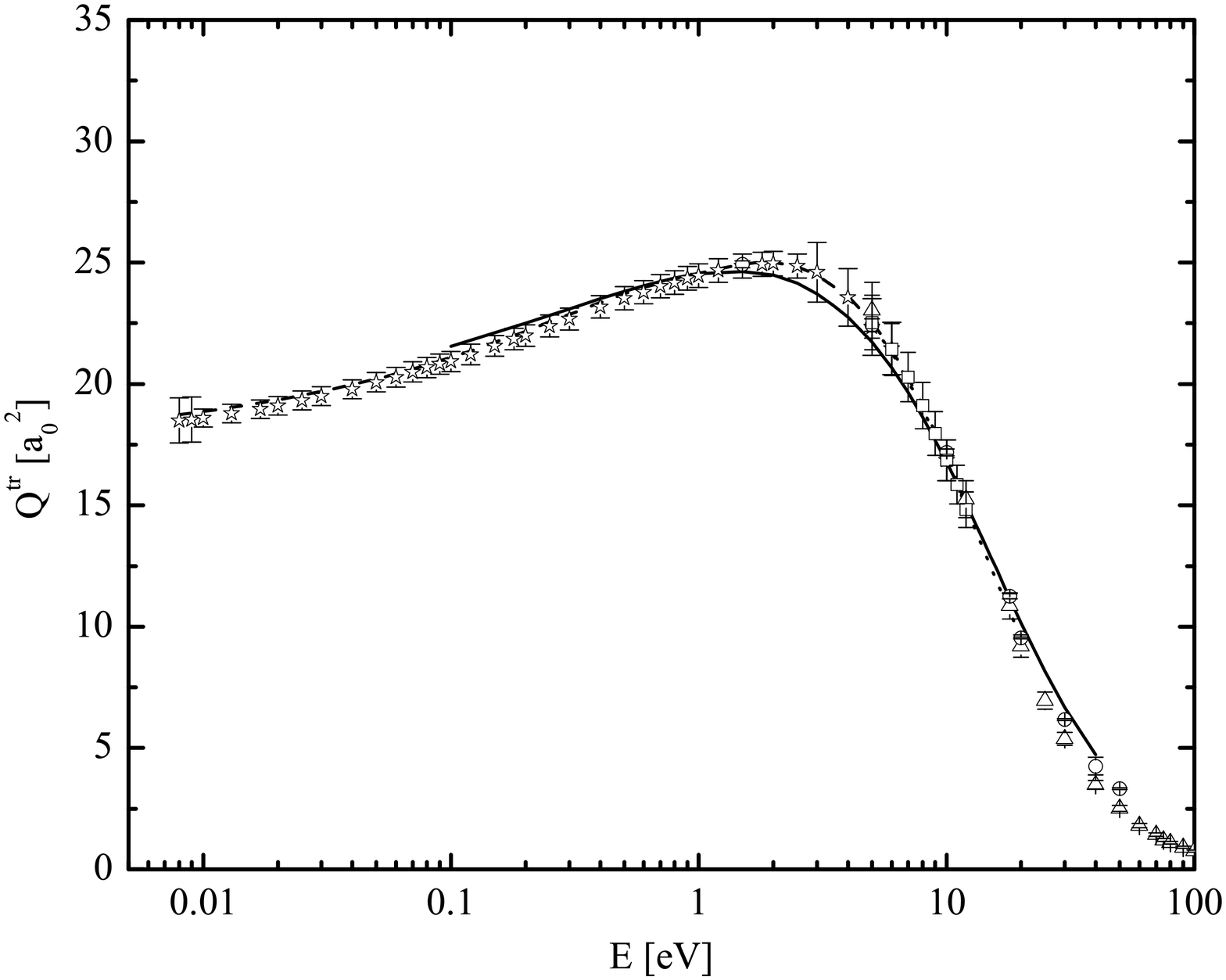}} \caption{Momentum transfer
cross section $Q^{tr}$ as a function of energy $E$. Calculated data
($\relbar$ line ) for helium together with results of other authors
\citet{nes79} ($\relbar$ $\relbar$ dashed line), \citet{fon81}
($\cdots$ dotted line), \citet{reg80} ($\bigtriangleup$), \citet{cro70}
($\star$), \citet{bru92} ($\circ$), \citet{mil77} ($\square$).}
\label{fig:Qtr}
\end{figure}

The equation (\ref{eq:RSE}) is solved by the partial wave method
described in \citet{ign97}. One obtains as the result the phase
shifts $\delta_{l}(E)$ for $l=0, 1,2,...$, where each
$\delta_{l}(E)$ corresponds to the partial wave with a given orbital
number $l$. This way of solving Eq.~(\ref{eq:RSE}) has advantages
since, apart from the transport cross-section
\begin{equation}
\label{eq:sigma_tr} Q_{tr}(E)=\frac{2\pi\hbar
c}{E}\sum\limits_{l=0}^{\infty}
(l+1)sin^{2}(\delta_{l}-\delta_{l+1}),
\end{equation}
it allows determination of the elastic cross-section, as well as
cross-sections of higher order (viscosity cross-section, etc.). As
one can see in Figure~\ref{fig:Qtr}, there is an excellent agreement
between the values of $Q_{tr}(E)$ calculated using the potential
(\ref{eq:U(r)}) and various experimental and theoretical data of
other authors.

\begin{table}
\caption{The parameter values of the model potential $U(r)$ from
Eq.~(\ref{eq:U(r)}) in atomic units}\label{tab:koef_U}
\begin{center}
\begin{tabular}{@{}|c|c|c|c|c|c|c|c|c|c|}
\hline
Z & $q$ & $a$ & $b$ & $c$ & $\alpha$ & $h$ & $r_{i}$ & $r_{f}$ & $r_{0}$ \\
\hline
2 & $0.7$ & $-0.59597652$ & $2.2545472$ & $-2.19405731$ & $1.384$ & $0.01$ & $0.73$ & $1.75$ & $0.9$\\
\hline
\end{tabular}
\end{center}
\end{table}

\section{Results and discussion}

\begin{figure}
\centerline{\includegraphics[width=\columnwidth,
height=0.75\columnwidth]{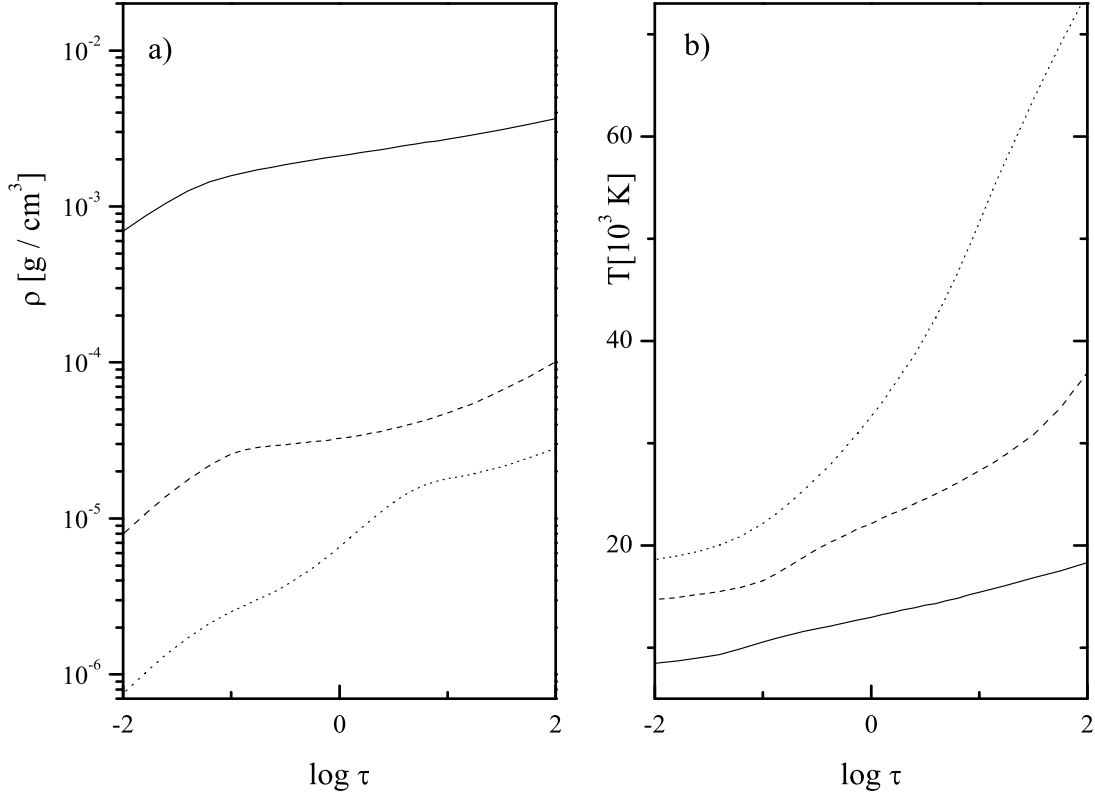}} \caption{DB white dwarf
atmosphere models with log $g=8$ and $T_{eff}$=12000K (full curve),
$T_{eff}$=20000K (dashed curve) and $T_{eff}$=30000K (dotted curve)
from \citet{koe80}: (a) The mass densities; (b) The temperatures, as
functions of Rosseland opacity $\tau$.} \label{fig:DBmodels}
\end{figure}

In order to apply our results to the study of DB white dwarf
atmosphere plasma properties, helium plasmas with electron ($N_{e}$)
and atom ($N_{a}$) densities and  temperatures ($T$), characteristic
of atmosphere models presented in the literature \citep{koe80}, are
considered here. So, the behaviour of $\rho$ and $T$ for models with
the logarithm of surface gravity $\log g=$8 and effective
temperature $T_{eff}=$12000K, 20000K and 30000K is shown in
Fig.~\ref{fig:DBmodels} as a function of Rosseland opacity $\tau$.
As one can see, these atmospheres contain layers of dense helium
plasma. In order to cover the considered plasma parameter range
reliably, we tested our method for calculation of the plasma
electrical conductivity within a wider range of mass density $1
\times 10^{-6} \textrm{g}/\textrm{cm}^{3} \lesssim \rho \lesssim 2
\textrm{g}/\textrm{cm}^{3}$ and temperature $1\cdot 10^{4}\textrm{K}
\lesssim T \lesssim 1\cdot 10^{5}\textrm{K}$.

\begin{figure}
\centerline{\includegraphics[width=\columnwidth,
height=0.75\columnwidth]{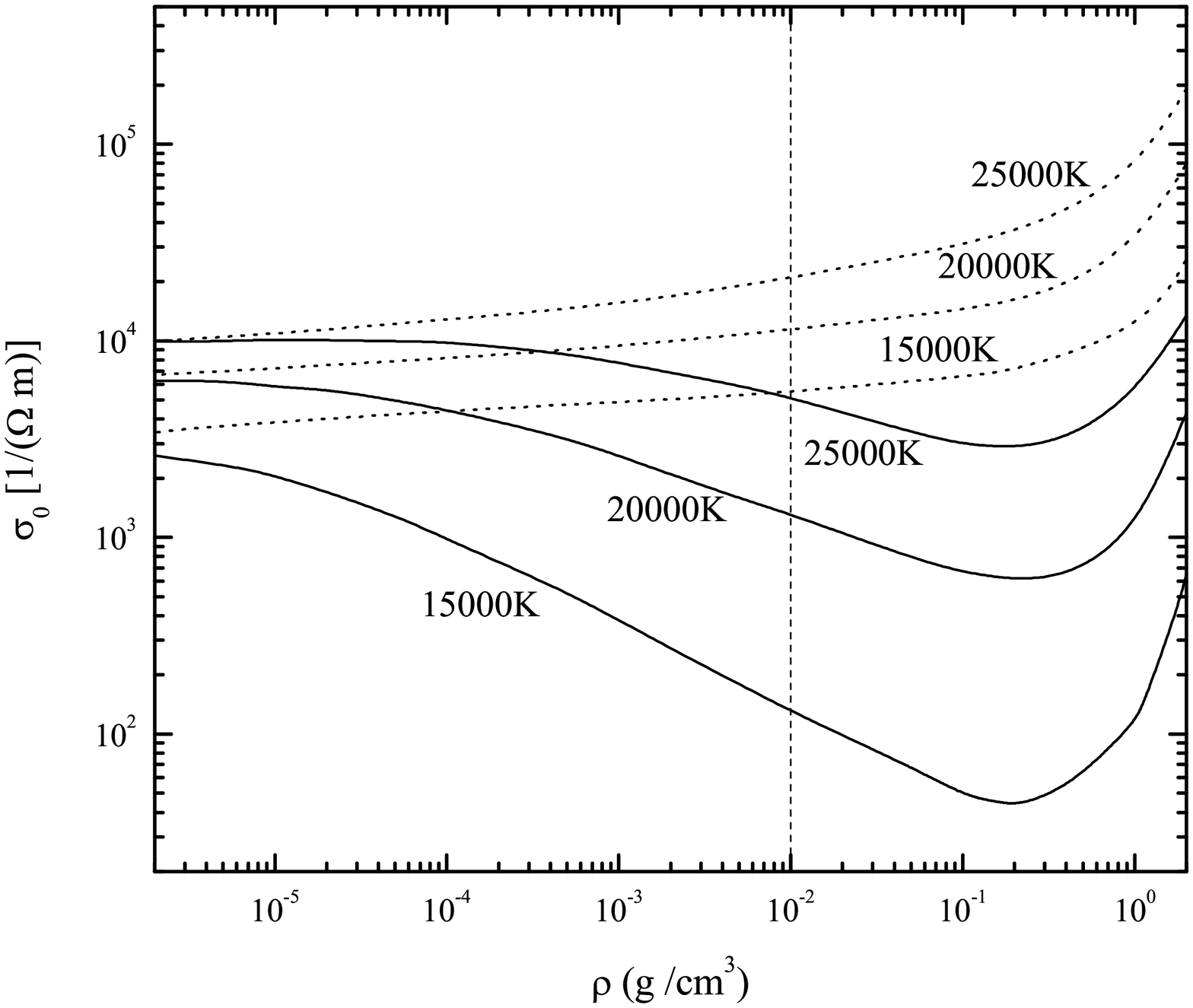}} \caption{Static electrical
conductivity $\sigma_{0}$ of dense He plasmas as a function of mass
density $\rho$ (full curves), compared to the Coulomb part of
conductivity (dashed curves). The area left to the vertical dashed
line marks the region which is of interest for DB white
dwarfs.\label{fig:withatoms}}
\end{figure}

The influence of neutral atoms on the electrical conductivity of
helium plasma is shown in Fig.~\ref{fig:withatoms}. In this figure
the electrical conductivities for $T=15000$, $20000$ and $25000$K
are given as functions of mass density $\rho$. The range between the
two vertical dashed lines corresponds to the conditions in the
considered DB white dwarf atmospheres.  Two groups of curves,
calculated using the expression (\ref{eq:sigma0}), are presented in
this figure: a) the dashed ones, obtained by neglecting the
influence of atoms, i.e. with $\nu_{ea}=0$; b) the full-line curves
calculated with the influence of atoms included, i.e. with
$\nu_{ea}$ given by Eq.~(\ref{eq:nu_ea}). First, one should note
that the behaviour of these two groups of curves is qualitatively
different: the first one increases constantly with the increase of
$\rho$, while the other group of curves decreases, reaches a
minimum, and then starts to increase with the increase of $\rho$.
One could explain such behaviour of the electrical conductivity by
the pressure ionization. This figure also clearly shows when the
considered plasma can be treated as "fully ionized".

\begin{figure}
\includegraphics[angle=0, width=\textwidth]{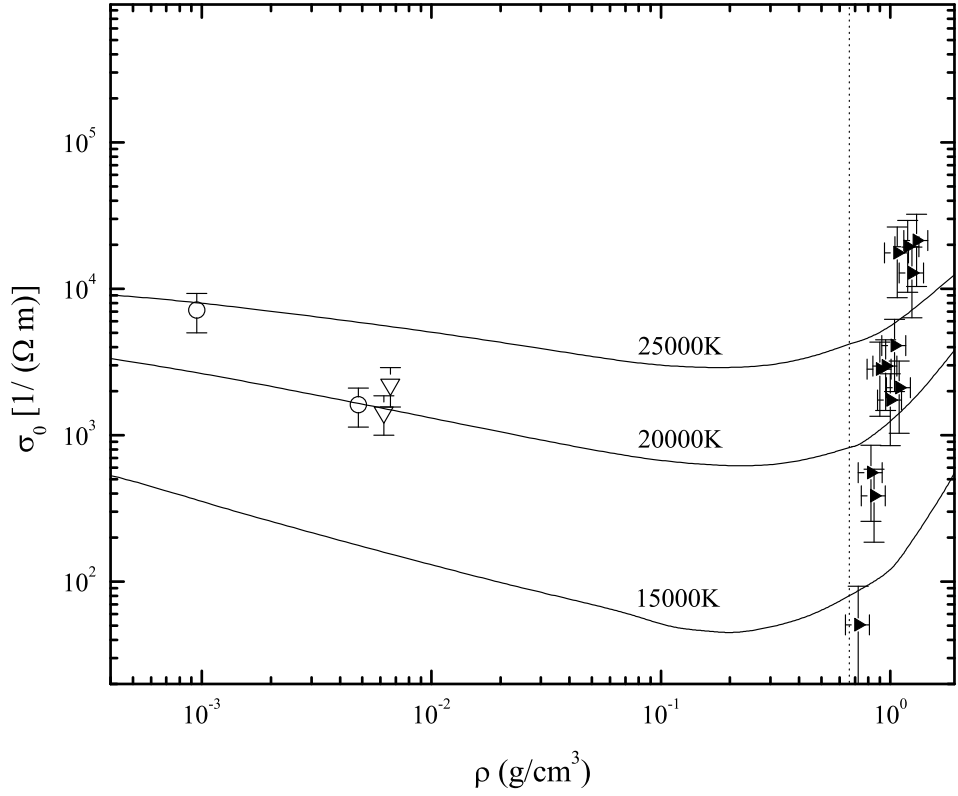}
\caption{Static electrical conductivity $\sigma_{0}$ of helium
plasma for various temperatures as a function of the mass density
$\rho$. Full line - the calculations based on the expressions
(\ref{eq:sigma0})-(\ref{eq:nu_ea}); right of the vertical dotted
line there is a region of extremely dense plasmas where one should
treat the presented calculation as an extrapolation.
$\bigtriangledown$ - \citet{shi03} 20000-23000K;
$\blacktriangleright$ - \citet{ter02}
 15000-25000K; $\bigcirc$ - \citet{min80} 20000-25000K} \label{fig:HeCond}
\end{figure}

In Fig.~\ref{fig:HeCond} we compare our values of the helium plasma
conductivity, shown by full curves for $T=15000$ K, 20000 K and
25000 K within the region $5\cdot 10^{-4}$g/cm$^{3}< \rho <
2$g/cm$^{3}$, with the existing experimental data. Let us note that
these experimental results are uniquely available for comparison so
that, in spite of their low accuracy, the agreement with them gives
the only possible indication that our results are correct.

Within the region $\rho < 0.65$ g/cm$^{3}$, i.e. to the left of the
vertical line in Fig.~\ref{fig:HeCond}, there are experimental
results from \citet{ter99} ($\circ$) and \citet{shi03}
($\triangledown$) where the temperature was determined with an error
of less than $20\%$, which are related to the temperature range
20000K-25000K. For $\rho > 0.65$ g/cm$^{3}$, i.e. right of the
vertical line in Fig.~\ref{fig:HeCond}, are shown several values of
the plasma conductivity, obtained by \citet{ter02} for the
temperature range 15000K-25000K. These experimental values are
obtained with an experimental error $\sim 50\%$ and can be treated
only as characteristic of this temperature region as a whole. These
data at least indicate that our results for $\rho > 0.65$ g/cm$^{3}$
lie in the correct domain of the electrical conductivity values.

\begin{figure}
\centerline{\includegraphics[width=\columnwidth,
height=0.75\columnwidth]{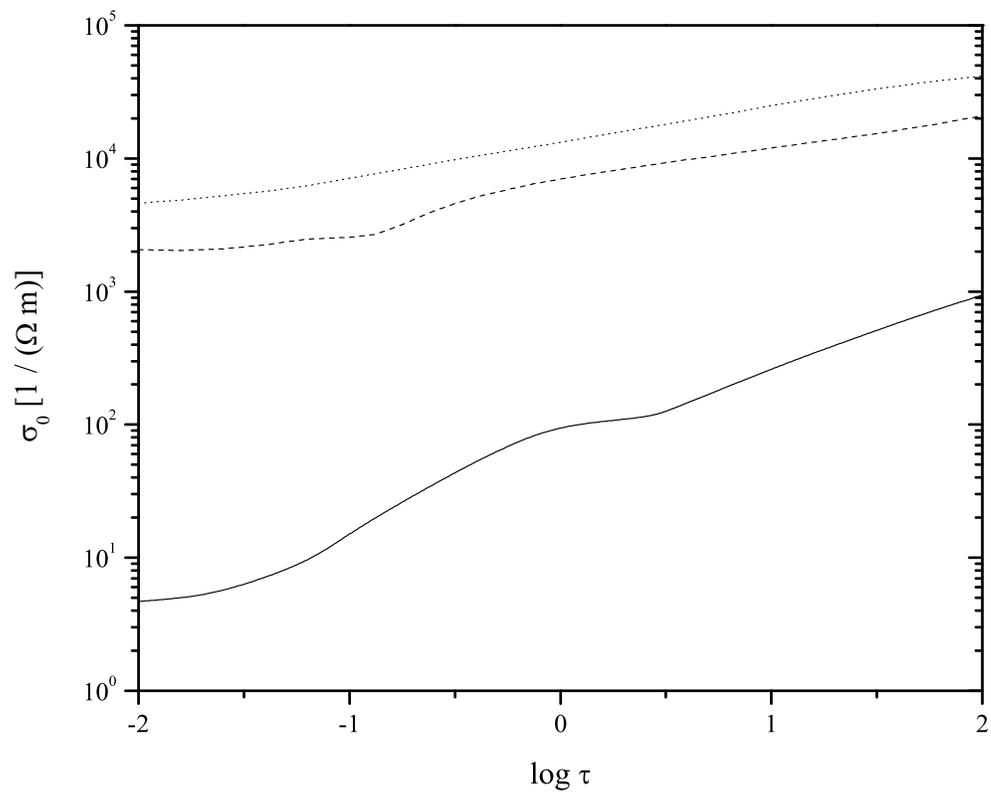}} \caption{Static electrical
conductivity $\sigma_{0}$ as a function of the logarithm of
Rosseland opacity $\tau$ for DB white dwarf atmosphere models with
log $g=8$ and $T_{eff}$=12000K (full curve), $T_{eff}$=20000K
(dashed curve) and $T_{eff}$=30000K (dotted curve)}
\label{fig:DBcond}
\end{figure}

The developed method was then applied to calculation of plasma
electrical conductivity for the models of DB white dwarf atmospheres
presented in Fig.~\ref{fig:DBmodels}. The results of the
calculations are shown in Fig.~\ref{fig:DBcond}. First, let us note
a regular behaviour of the static electrical conductivity which one
should expect considering the characteristics of DB white dwarf
atmospheres. Further, the electrical conductivity profiles presented
in this figure show that, for the considered DB white dwarf models,
plasma electrical conductivity changes over the domain of values
where our results agree with the experimental ones (see
Fig.~\ref{fig:HeCond}). This indicates that the theoretical
apparatus presented here may be adequate to be used for
investigation of DB white dwarfs in the magnetic field of their
partners in binary systems and magnetic white dwarfs.

\begin{table}
\caption{Static electrical conductivity of helium plasma
$\sigma_{0}[1/(\Omega m)]$}
\begin{tabular}{@{} r c c c c c c c c c @{} } \hline\hline
\multicolumn{1}{c}{} & \multicolumn{9}{c}{$\rho [g/cm^{3}]$}
\\ \cline{2-10}
$T[K]$&5.00E-07 &1.00E-06 &5.00E-06 &1.00E-05 &5.00E-05 &1.00E-04 &5.00E-04 &1.00E-03 & 5.00E-03 \\
\hline\hline
 8000 & 6.53E+00& 4.69E+00& 2.15E+00& 1.50E+00& 6.90E-01& 4.89E-01& 2.20E-01& 1.53E-01& 6.88E-02 \\
 9000 & 4.34E+01& 3.16E+01& 1.47E+01& 1.13E+01& 5.12E+00& 3.60E+00& 1.58E+00& 1.19E+00& 5.20E-01 \\
10000 & 1.69E+02& 1.30E+02& 6.74E+01& 5.02E+01& 2.38E+01& 1.72E+01& 8.12E+00& 5.73E+00& 2.56E+00 \\
12000 & 9.36E+02& 8.04E+02& 5.09E+02& 4.05E+02& 2.19E+02& 1.66E+02& 8.35E+01& 6.08E+01& 2.92E+01 \\
14000 & 2.27E+03& 2.08E+03& 1.62E+03& 1.38E+03& 8.87E+02& 6.97E+02& 3.99E+02& 3.00E+02& 1.28E+02 \\
16000 & 3.64E+03& 3.55E+03& 3.15E+03& 2.87E+03& 1.45E+03& 1.18E+03& 6.85E+02& 5.28E+02& 2.75E+02 \\
18000 & 4.90E+03& 4.97E+03& 4.76E+03& 3.79E+03& 2.87E+03& 2.46E+03& 1.58E+03& 1.26E+03& 7.07E+02 \\
20000 & 6.08E+03& 6.22E+03& 6.32E+03& 5.61E+03& 4.99E+03& 4.46E+03& 3.21E+03& 2.53E+03& 1.58E+03 \\
25000 &         &         & 9.95E+03& 1.02E+04& 9.90E+03& 9.90E+03& 8.62E+03& 7.74E+03& 5.95E+03 \\
30000 &         &         & 1.33E+04& 1.40E+04& 1.56E+04& 1.60E+04& 1.57E+04& 1.50E+04& 1.21E+04 \\
35000 &         &         & 1.65E+04& 1.74E+04& 2.02E+04& 2.15E+04& 2.34E+04& 2.35E+04& 2.15E+04 \\
40000 &         &         & 1.97E+04& 2.09E+04& 2.43E+04& 2.61E+04& 3.06E+04& 3.18E+04& 3.22E+04 \\
45000 &         &         & 2.31E+04& 2.45E+04& 2.84E+04& 3.06E+04& 3.69E+04& 3.93E+04& 4.36E+04 \\
55000 &         &         & 3.04E+04& 3.23E+04& 3.70E+04& 3.98E+04& 4.81E+04& 5.26E+04& 6.40E+04 \\
65000 &         &         & 3.81E+04& 4.08E+04& 4.65E+04& 4.98E+04& 6.00E+04& 6.56E+04& 8.20E+04 \\
75000 &         &         & 4.58E+04& 4.93E+04& 5.66E+04& 6.06E+04& 7.26E+04& 7.92E+04& 9.84E+04 \\
\hline\hline
\end{tabular}
\label{tab:ec}
\end{table}

In order to provide possibility for direct applications of our
results to different theoretical investigations the values of static
electrical conductivity $\sigma_{0}$ of helium plasma in a wide
range of $\rho$ and $T$ are given in Table~\ref{tab:ec}. This table
covers plasma conditions for all models of DB white dwarf
atmospheres presented in \citet{koe80} and the corresponding values
of $\sigma_{0}$ were determined for the helium plasmas in the state
of local thermodynamical equilibrium with given $\rho$ and $T$.

The method developed in this paper represents also a powerful tool
for research into white dwarfs with different atmospheric
compositions (DA, DC etc.), and for investigation of some other
stars (M-type red dwarfs, Sun etc.). Finally, the presented method
provides a basis for the development of methods to describe other
transport characteristics which are important for the study of all
mentioned astrophysical objects, such as the electronic
thermo-conductivity in the star atmosphere layers with large
electron density, electrical conductivity in the presence of strong
magnetic fields and dynamic (high frequency) electrical
conductivity.

\section*{Acknowledgments}

This work was supported by the Ministry of Science and Technological
Development of Serbia as a part of the project "Radiation and
transport properties of the non-ideal laboratory and ionospheric
plasma" (Project number 141033) and "Influence of collisional
processes on astrophysical plasma line shapes" (Project number
146001).



\newpage


\end{document}